\documentclass[journal,twoside,web]{ieeecolor}
\usepackage{cite, amsmath, graphicx, threeparttable, xcolor, float, multirow, tikz, pifont, makecell, booktabs}
\usepackage[hidelinks]{hyperref}
\usepackage[outdir=./]{epstopdf}

\def\logoname{LOGO-generic-web}
\definecolor{subsectioncolor}{rgb}{0,0.541,0.855}
\def\journalname{arXiv}

\def\BibTeX{{\rm B\kern-.05em{\sc i\kern-.025em b}\kern-.08em
    T\kern-.1667em\lower.7ex\hbox{E}\kern-.125emX}}
\DeclareRobustCommand{\orcidicon}{
\begin{tikzpicture}
\draw[lime, fill=lime] (0,0)
circle[radius=0.16]
node[white]{{\fontfamily{qag}\selectfont \tiny \.{I}D}};
\end{tikzpicture}
\hspace{-2mm}
}
\foreach \x in {A, ..., Z}{
\expandafter\xdef\csname orcid\x\endcsname{\noexpand\href{https://orcid.org/\csname orcidauthor\x\endcsname}{\noexpand\orcidicon}}
}

\begin{document}
\title{A Deep Registration Method for Accurate Quantification of Joint Space Narrowing Progression in Rheumatoid Arthritis}
\author{Haolin Wang\hspace{-1.5mm}\orcidW{}, Yafei Ou\hspace{-1.5mm}\orcidO{}, Wanxuan Fang\hspace{-1.5mm}\orcidF{}, Prasoon Ambalathankandy\hspace{-1.5mm}\orcidP{}, \IEEEmembership{Member, IEEE}, Naoto Goto, Gen Ota, Masayuki Ikebe\hspace{-1.5mm}\orcidI{}, \IEEEmembership{Member, IEEE} and Tamotsu Kamishima\hspace{-1.5mm}\orcidK{}
\thanks{Haolin Wang and Yafei Ou contributed equally. \textit{(Corresponding author: Yafei Ou.)}}
\thanks{Haolin Wang and Wanxuan Fang are with the Graduate School of Health Sciences, Hokkaido University, Sapporo 060-0812, Japan (e-mail: haolin.wang.k3@elms.hokudai.ac.jp; wanxuan.fang.z7@elms.hokudai.ac.jp).}
\thanks{Yafei Ou, Prasoon Ambalathankandy, Naoto Goto, Gen Ota and Masayuki Ikebe are with the Research Center For Integrated Quantum Electronics, Hokkaido University, Sapporo 060-0813, Japan, and also with the Graduate School of Information Science and Technology, Hokkaido University, Sapporo 060-0814, Japan (e-mail: yafei.ou.x5@elms.hokudai.ac.jp; prasoon.ak@ist.hokudai.ac.jp; lion110king44@eis.hokudai.ac.jp; genota83@eis.hokudai.ac.jp; ikebe@ist.hokudai.ac.jp).}
\thanks{Tamotsu Kamishima is with Faculty of Health Sciences, Hokkaido University, Sapporo 060-0812, Japan. (e-mail: ktamotamo2@hs.hokudai.ac.jp).}
}

\maketitle

\begin{abstract}
Rheumatoid arthritis (RA) is a chronic autoimmune inflammatory disease that results in progressive articular destruction and severe disability.
Joint space narrowing (JSN) progression has been regarded as an important indicator for RA progression and has received sustained attention.
In the diagnosis and monitoring of RA, radiology plays a crucial role to monitor joint space. A new framework for monitoring joint space by quantifying JSN progression through image registration in radiographic images has been developed. This framework offers the advantage of high accuracy, however, challenges do exist in reducing mismatches and improving reliability. 
In this work, a deep intra-subject rigid registration network is proposed to automatically quantify JSN progression in the early stage of RA. 
In our experiments, the mean-square error of Euclidean distance between moving and fixed image is 0.0031, standard deviation is 0.0661 mm, and the mismatching rate is 0.48\%.
The proposed method has sub-pixel level accuracy, exceeding manual measurements by far, and is equipped with immune to noise, rotation, and scaling of joints.
Moreover, this work provides loss visualization, which can aid radiologists and rheumatologists in assessing quantification reliability, with important implications for possible future clinical applications.
As a result, we are optimistic that this proposed work will make a significant contribution to the automatic quantification of JSN progression in RA.

\end{abstract}

\section{Introduction}
Rheumatoid arthritis (RA) is a chronic autoimmune inflammatory disease marked by joint swelling and tenderness, leading to progressive articular destruction combined with severe disability. Joint space narrowing (JSN) due to destroyed cartilage may have a more significant effect on functional status than erosions, making it a valid case for treatment \cite{pfeil2013usefulness}. Early diagnosis and treatment with disease-modifying antirheumatic drugs (DMARDs) therapy can prevent irreversible disability by arresting RA before irreversible damage is done to the joints, thereby avoiding or significantly slowing the progression of joint damage in 90\% of patients \cite{platten2017fully}.  Therefore, using inexpensive, convenient, widely available imaging techniques with high sensitivity and specificity is essential for early diagnosis of RA and early intervention and management \cite{aletaha2018diagnosis}.

Radiography has proven effective in identifying RA patients at a higher risk of further damage progression, and detecting early joint damage through radiography holds significant prognostic value. Sharp/van der Heijde scoring method (SvdH) \cite{van2000read}, the current gold standard for assessing radiological progression in clinical practice, provides a method for scoring JSN and erosion of the hands/wrists and feet \cite{rydell2021predictors}. However, this approach is time-consuming and prone to significant variability among radiologists and rheumatologists, even after professional training \cite{minh2022application}. Therefore, to minimize the above-mentioned weaknesses, many recent studies have been devoted to automatically quantifying the joint space of RA \cite{peloschek2007automatic, langs2008automatic, huo2015automatic, hirano2019development, ou2022sub}.

\subsubsection{Medical image analysis in joint space of RA}

According to the nature of the methods and their output metrics, previous works on joint space quantification for RA can be divided into three frameworks, edge detection based joint space width (JSW) quantification framework, classification based SvdH scoring framework, and registration based JSN progression quantification framework. As shown in Table \ref{tab:feature}, each of these three frameworks have their advantages and disadvantages under different evaluation standards.

\begin{table}[!t]
\caption{Feature comparison of mainstream frameworks of joint space quantification in RA.}
\label{tab:feature}
\centering
\setlength{\tabcolsep}{1mm}{
\begin{tabular}{cccccc}
\toprule
&Method&Output&Sensitivity&Detectable&Purpose \\
\midrule
\cite{langs2008automatic}&Margin detection&JSW&Medium&Early stage&Both \\
\cite{hirano2019development}&Classification&SvdH score&Low&All stages&Qualitative \\
\cite{ou2022sub}&Registration&JSN&High&Early stage&Quantitative \\
\bottomrule
\end{tabular}
}
\end{table}

The margin detection based JSW quantification framework is the earliest computer-aided methodology in RA, it can be performed as follow: (i) Detect bone margin by using supervised marchine learning (ML) network \cite{langs2008automatic} or image feature \cite{huo2015automatic} such as intensity, gradient and derivative. (ii) Fit polynomial functions to bone margin curves. (iii) Quantify JSW according to the distance between polynomial functions. This framework has a wide application prospect. According to the quantified absolute JSW, the SvdH scores can be scored for qualitative diagnosis, and on the other hand, the JSN progression (relative JSW) can be calculated for quantitative monitoring. Nevertheless, it has some limitations: (i) Since this framework relies on margin information to determine the JSW, it is only suitable for use in the early stages of RA when there is a clear bone margin. (ii) This framework only can achieve pixel-level accuracy, as it limits the sensitivity of joint space monitoring.

To overcome these limitations, a ML classification based SvdH scoring framework was proposed for rapid diagnosis in any stage of progression. In this framework, a supervised trained classifier is used to classify hand joint images from level 0 to level 4, such as CNN \cite{hirano2019development} or SVM \cite{nakatsu2020finger}. This framework can be used in any stage of RA, and can help rheumatologist make qualitative assessment. 
Due to the inherent characteristics of classification, this framework is not suitable for quantitative analysis and has low sensitivity.

\begin{figure}[!t]
\centering\includegraphics[width=\linewidth]{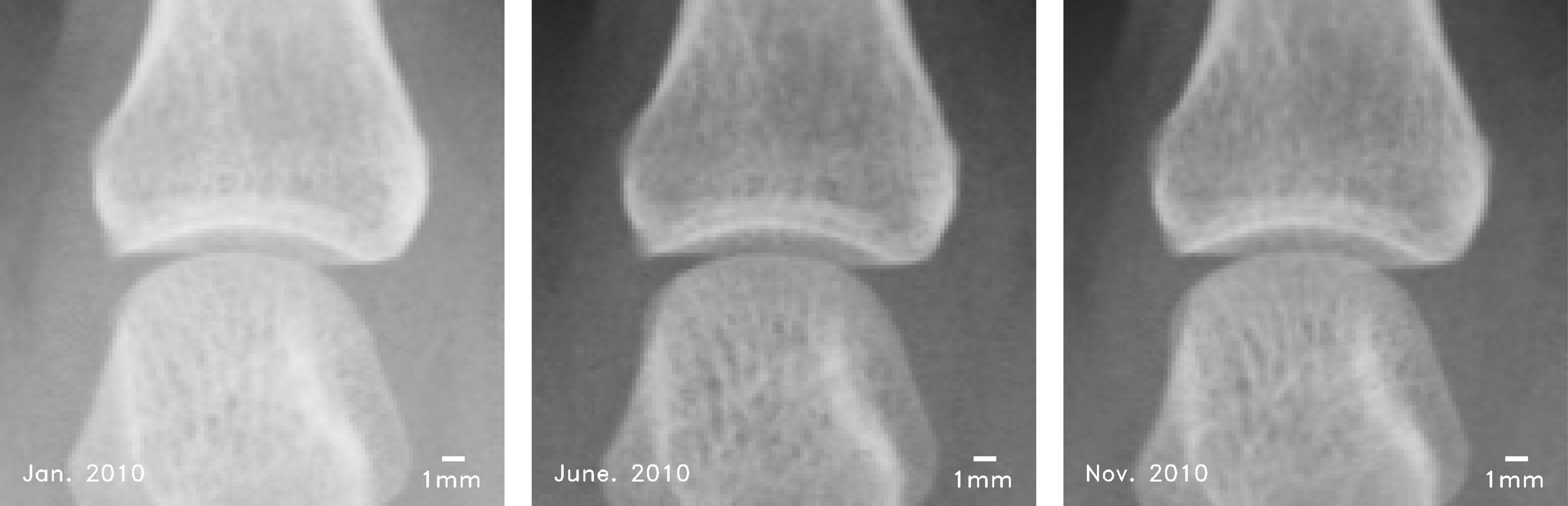}
\caption{JSN progression of a MCP joint for little finger over a period of 10 months. From left to right the images are: baseline, five-month, and ten-month images (spatial resolution: 0.175 mm/pixel). Usually, JSN progression is less than one pixel per year, therefore, it is difficult for radiologists and rheumatologists to see. Operating with an algorithm with pixel level accuracy to quantify JSN progression over a period of one year can be ineffective. JSN progression measured for five and ten months X-rays relative to baseline using this work are -0.197 pixel and 0.174 pixel respectively.}
\label{fig:JSN_ex}
\end{figure}

JSN progression is an important indicator for drug management in RA and has received widespread attention.
However, over the course of one year, JSN progression can be less than one pixel, making it difficult to detect, as show in Fig.~\ref{fig:JSN_ex}. 
To quantitatively monitor JSN progression with high accuracy, a registration based JSN progression quantification framework was proposed \cite{ou2022sub, ou2019automatic}.
Take a metacarpophalangeal (MCP) joint as an example, this framework can be performed as follows: (i) Segment the proximal phalanx bone and metacarpal bone. (ii) Measure the displacements of the proximal phalanx bone and metacarpal bone between a baseline and its follow-up finger joint images respectively by using image registration algorithm. (iii) Calculate the displacement difference between the proximal phalanx bone and metacarpal bone to measure JSN progression.

Compared to other frameworks, this framework has potential for higher sensitivity and lower mean error.
However, this framework also has some limitations: (i) Changes of bone characteristics caused by bone erosion in advanced RA can reduce the accuracy of rigid registration algorithm and even cause mismatching. For the above reason, this framework is mostly used in early stage of RA. (ii) Considering that registration algorithm only can provide JSN progression, this limits its application for qualitative diagnosis.

\begin{figure*}[!t]
\includegraphics[width=\textwidth]{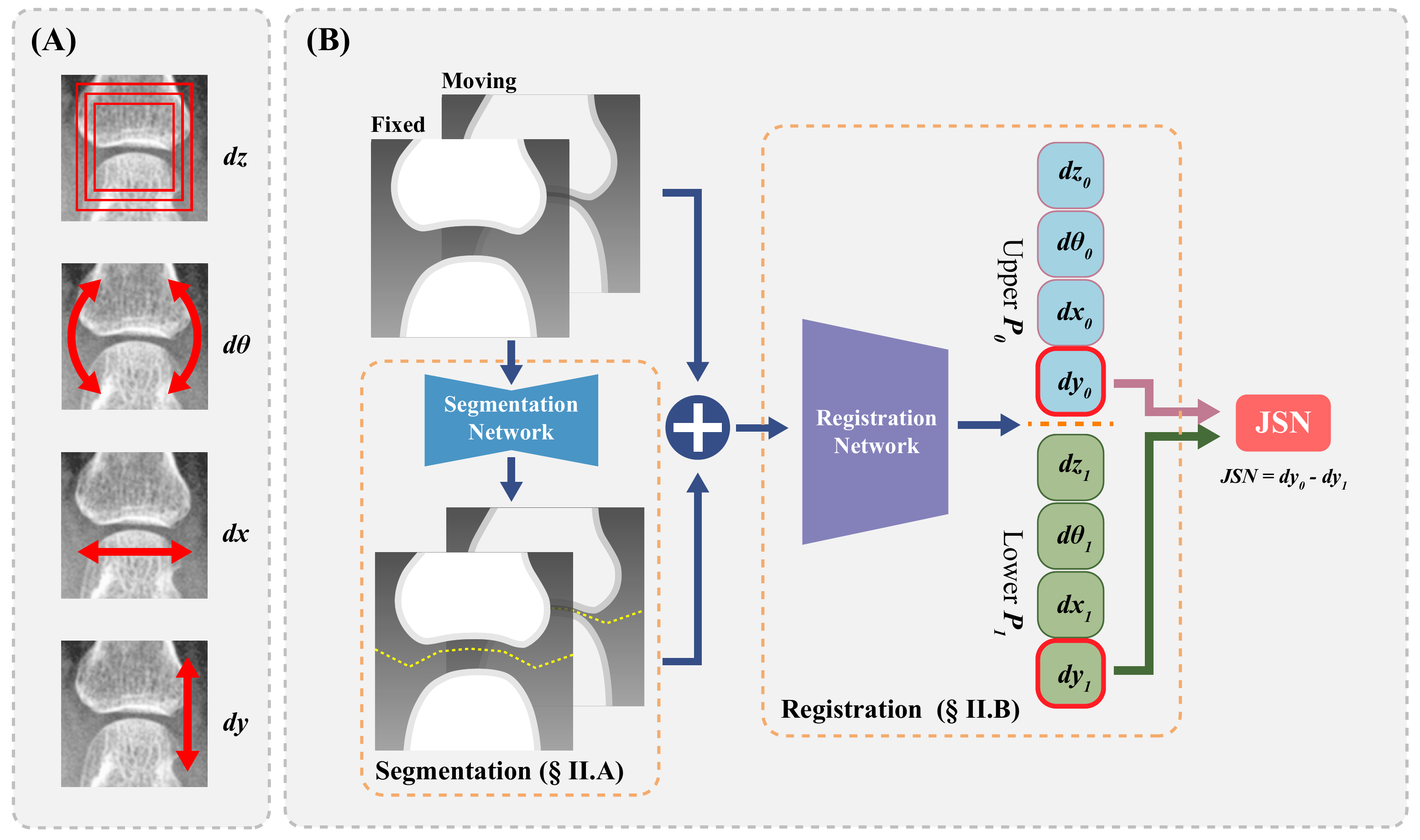}
\caption{(A) Four rigid transformation parameters are shown that are used in this work; $dz$: scaling, $d\theta$: rotation, $dx$: displacement on x-axis, $dy$: displacement on y-axis. (B) The overview of our proposed deep learning image registration based JSN progression quantification methodology. 
This work can be divided into two steps: joint segmentation, and JSN progression quantization.
Take a MCP joint as an example, this work can be performed as follow: (i) \textsection~\ref{sec:segmentation} A supervised U-net++ based network is implemented to segment the proximal phalanx bone and metacarpal bone region of the MCP joint. (ii) \textsection~\ref{sec:JSNQuanti} An un-supervised ResNet-like based deep registration network is proposed to quantify the rigid transformation parameters of the proximal phalanx bone and metacarpal bone region. (iii) The JSN progression can be obtained by calculating the displacement difference on y-axis between two bone region.}
\label{fig:teaser}
\end{figure*}

\subsubsection{ML-based image registration in medical image analysis}
Registration in medical image processing refers to the process of aligning multiple medical images on a common coordinate system with matched contents, and it is also an important step in many medical image analysis tasks \cite{chen2022recent}.
At present, the development of ML-based medical image registration algorithms has been gaining popularity \cite{chen2021deep}.
According to the deformation model type, registration algorithms can be divided into rigid and non-rigid (deformable) registration.
Non-rigid registration has broad applications in tissues or organs such as the brain, thorax, lung \cite{fu2020deep}, and can be used to detect lung motion \cite{ehrhardt2010statistical} and tumor regression \cite{neylon2017neural}.
For rigid registration, the transformation parameters are obtained based on the convolutional neural network to achieve image registration. A 2D/3D target registration in X-ray image using convolutional neural networks was presented to obtain transformation parameters \cite{miao2016cnn}.
With this inspiration, our work focuses on the overall movement of bones, thus, the rigid registration algorithm can achieve higher quantification accuracy.
By utilizing the potential of deep learning, the problems of the traditional computation-based image registration methods mentioned in the previous subsection can be addressed.
This work is limited to detecting JSN progression of the same patient over consecutive time points \cite{stoel2020use}, that is, intra-subject registration.

\subsubsection{Our contributions}
In this work, a deep learning-based methodology is proposed for JSN quantification, and the following are our original contribution:
\begin{enumerate}
\item Implemented an image segmentation network based on U-net++ to segment joint images.
\item Proposed a ResNet-like deep registration network to measure bone displacement.
\item The proposed method achieves sub-pixel accuracy monitoring of joint space in the early stage of RA.
\item Compared to related works, this work significantly improves robustness for scaling, rotation, and noise. This reduces mismatching caused by inconsistent angles between the upper and lower bones of the joint, variable spatial resolutions of radiography images, and inconsistent projection angles.
\item Our method provides a visualization loss that enables radiologists and rheumatologists to assess the reliability of quantification. This has important implications for the future clinical application of our method.
\end{enumerate}

The rest of this paper is organized as follows. \textsection~\ref{sec:method} describes the implementation of our methodology; including joint segmentation network and JSN progression quantification network, and introduces the clinical datasets used in this work. \textsection~\ref{sec:exp} presents and discuss the segmentation and registration results. \textsection~\ref{sec:conl} concludes this work, and discusses possible future research directions for computer-aided monitoring of RA.

\section{Methodology and materials}
\label{sec:method}
In this work, a deep learning based JSN quantification method is proposed. The proposed method can improve the sensitivity, accuracy and robustness of JSN progression monitoring in early stage of RA. As shown in Fig.~\ref{fig:teaser}, this work contains two networks: an Unet++ based joint segmentation network and a ResNet-like deep registration network for JSN progression quantification.

\subsection{Joint segmentation}
\label{sec:segmentation}
\begin{figure}[!t]
\includegraphics[width=\linewidth]{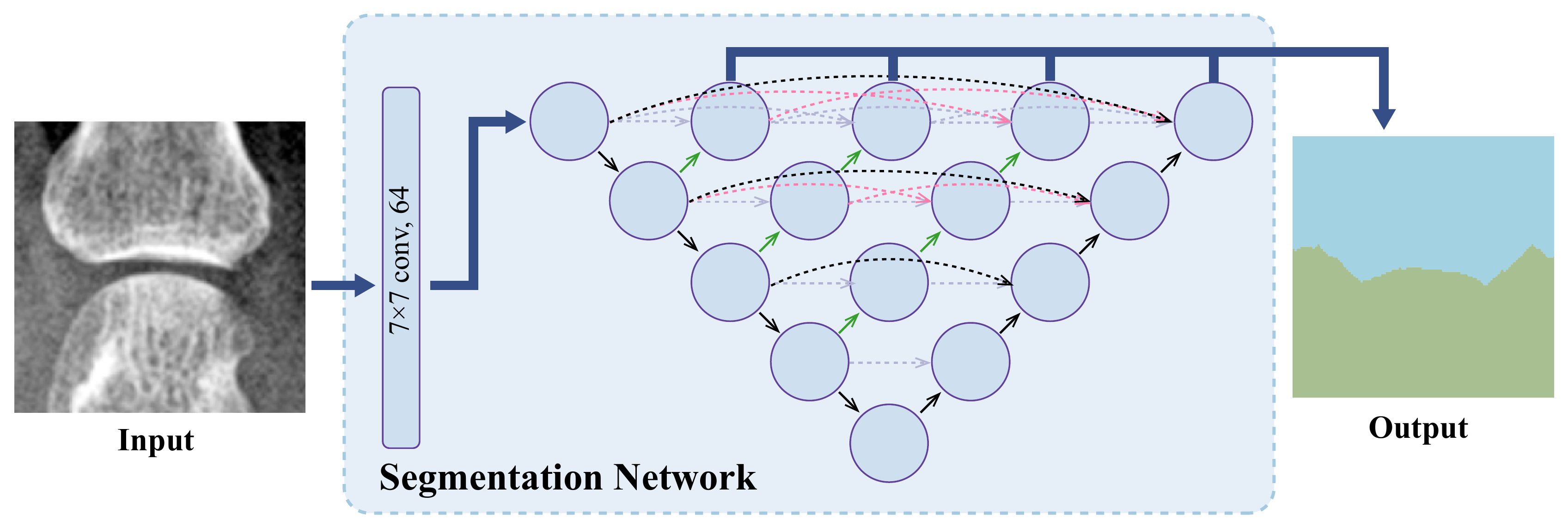}
\caption{The diagram of our segmentation network. This segmentation network contains one convolutional layer (kernel size: $7\times7$, channels: 64) and a 5-layer Unet++ network.}
\label{fig:segeNet}
\end{figure}

A network based on U-net++ with an added convolution layer is proposed for joint segmentation, as illustrated in Fig.~\ref{fig:segeNet}. Take a MCP joint as an example, the proximal phalanx bone and metacarpal bone are segmented separately using the U-net++ network. 
Thus, the displacement of the proximal phalanx bone (upper part of joint) and metacarpal bone (lower part of joint) can be measured separately. The output of the segmentation network is defined as $S$, where 0 represents the metacarpal bone region (upper region) and 1 represents the proximal phalanx bone region (lower region).

\begin{figure*}[!t]
\includegraphics[width=\textwidth]{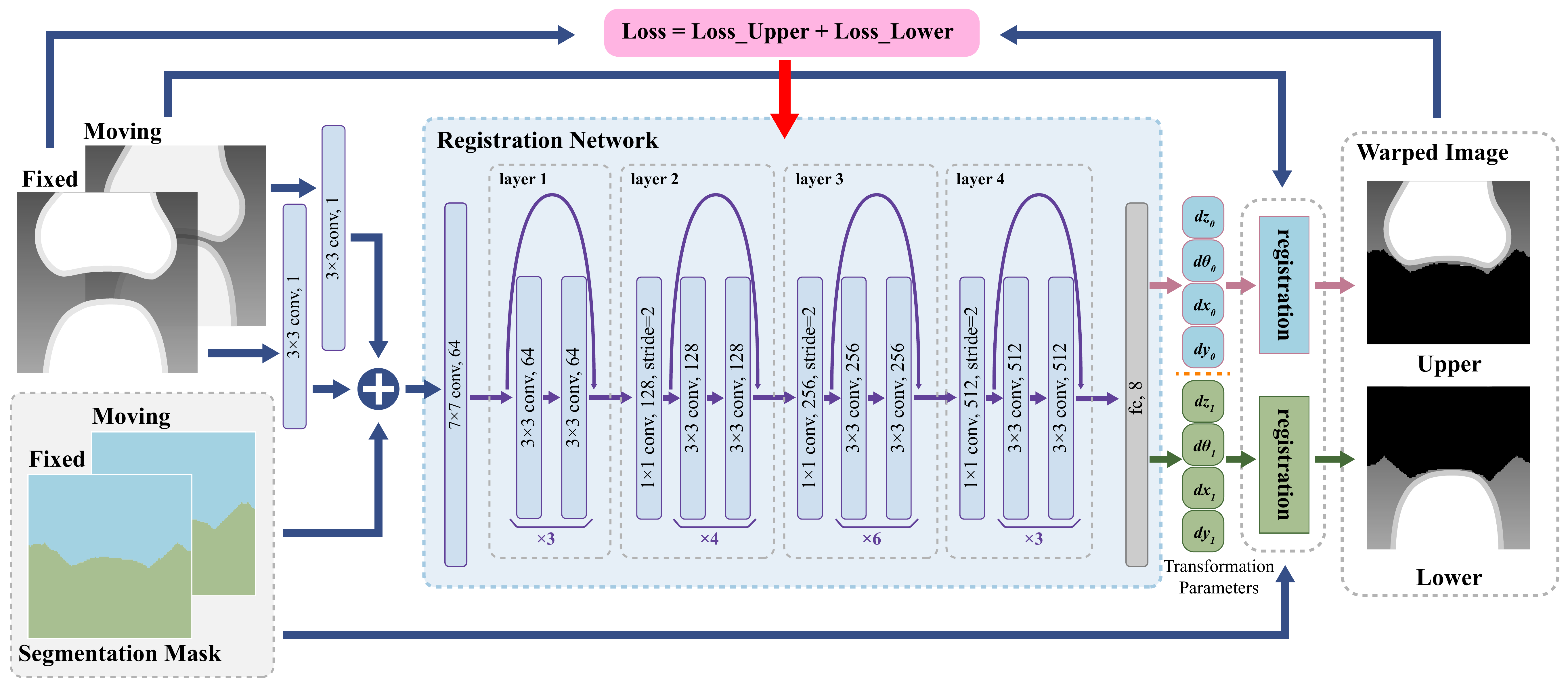}
\caption{The structure diagram of image registration network. In this case, after a convolution layer of $3\times3$ convolution kernels and a 1-channel convolution base and combined with its corresponding segmentation mask, the input image set is input into the registration network. The registration network contains a layer of convolution and 4 layers of residual convolution modules with the channels of 64, 128, 256, 512 respectively. The final transformation parameters are obtained after a full connection layer with 8 channels as output. These transformation parameters are used to deform the moving image to generate a warped image, and the difference between the generated warped image and the fixed image is defined as the loss, which is used to optimize the registration network.}
\label{fig:regiNet}
\end{figure*}

\subsection{JSN quantification by image registration}
\label{sec:JSNQuanti}
In this subsection, an unsupervised intra-subject rigid registration network is proposed for quantifying JSN progression. The pipeline can be explained as follows: (i) Quantify the transformation parameters between the baseline and follow-up radiographic images using the registration network, as shown in Fig.~\ref{fig:regiNet}. (ii) Calculate JSN progression based on the vertical displacement difference between the upper and lower regions of the joint.

\subsubsection{Transformation Parameterization}
A 3D rigid transformation can be parametrized by three in-plane and one out-of-plane transformation parameters \cite{kaiser20142d}, as shown in Fig.~\ref{fig:teaser}~(A).
The in-plane transformation parameters include two displacement parameters $dx$, $dy$ and one rotation parameter $d\theta$.
The out-of-plane transformation parameter is the scaling parameter $dz$.
In our registration network, the upper and lower regions are registered simultaneously, and two sets of parameters are introduced, namely:
 $\left \{ P_0 \mid dz_0, d\theta_0, dx_0, dy_0  \right \}$ for upper region, and $\left \{ P_1 \mid dz_1, d\theta_1, dx_1, dy_1 \right \}$ for lower region. Among them, the vertical displacement parameter $dy$ is used to calculate the JSN progression.

\subsubsection{Registration network}
The rigid transformation parameters of the upper and lower regions are obtained by simultaneously registering both regions. The detailed operation are described as follows.

Given a fixed joint image $F$ and a moving joint image $G$, they can be divided into upper and lower regions. The segmentation mask is denoted as $S$, where 0 represents the upper bone region and 1 represents the lower bone region.

The transformation matrix of the upper region is denoted as $t_0$ and the lower region is denoted as $t_1$, are generated based on the parameter sets $P_0$ or $P_1$ obtained through the proposed network, which is defined as in Eq.~\ref{eq:t}. 
\begin{equation}
\begin{split}
t\!=\!\!\begin{pmatrix}
dzcosd\theta & \!-dzsind\theta\! & \!dxdzcosd\theta\!-\!dydzsind\theta \\
dzsind\theta & dzcosd\theta & \!dxdzsind\theta\!+\!dydzcosd\theta \\
0 & 0 & 1
\end{pmatrix} 
\end{split}
\label{eq:t}
\end{equation}

Subsequently, the transformation matrix $t_0$ and $t_1$ is applied for the transformation function $T$ of the moving image $G$ to transform the spatial position of each pixel and generate the warped image $G_{t_0}$, $G_{t_1}$, which can be defined as shown in Eq.~\ref{eq:transform}:
\begin{equation}
\begin{split}
G_{t_0} = T\left (G_0, t_0 \right ) \quad
G_{t_1} = T\left (G_1, t_1 \right )
\end{split}
\label{eq:transform}
\end{equation}

Thus, the upper region $F_0$ and lower region $F_1$ of the fixed image, the upper region $G_0^{\prime}$ and lower region $G_1^{\prime}$ of the warped image can be expressed as follow:
\begin{equation}
\begin{split}
F_0=F*\neg S \quad F_1=F*S \\
G_0^{\prime}=G_{t_0}*\neg S \quad G_1^{\prime}=G_{t_1}*S
\end{split}
\label{eq:transSegmen}
\end{equation}

Then, the warped image $G^{\prime}$ can be obtained according to its upper region $G_0^{\prime}$ and lower region $G_1^{\prime}$, as shown in Eq.~\ref{eq:warped}.
\begin{equation}
\begin{split}
G^{\prime}=G_0^{\prime}+G_1^{\prime}
\end{split}
\label{eq:warped}
\end{equation}

In this registration network, the mean squared error (MSE) of the Euclidean distance is determined as the loss.
It can be determined as in Eq.~\ref{eq:loss}. Here, the $m$ and $n$ denote the width and hight of $F(x, y)$ respectively.
\begin{equation}
L(F, G) = \sqrt{\frac{1}{m\times n}\sum_{y=1}^{n}\sum_{x=1}^{m}(F(x, y)-G(x, y))^{2}}
\label{eq:loss}
\end{equation}

Here, $L(F, G)$ represents the Euclidean distance loss between image $F$ and image $G$.
The loss in our registration network includes both the upper and lower parts. For example, given a fixed image $F$ and a warped image $G^{\prime}$, the warped loss $L(F, G^{\prime})$ can be defined as follows:
\begin{equation}
L(F, G^{\prime}) = \alpha \times L(F_0, G_0^{\prime})+\beta \times L(F_1, G_1^{\prime})
\label{eq:loss2}
\end{equation}

$\alpha$ and $\beta$ represent the weights used to balance the loss of the upper and lower parts of the joint and it is set to $\alpha = \beta = 0.5$. Then, the original loss $L(F, G)$ of the moving image $G$ can be similarly calculated.

The transformation parameters $P_0$ and $P_1$ obtained from registering the upper and lower joint regions are used to generate the final results. The vertical displacements $dy_0$ and $dy_1$ are used to calculate JSN progression. Thus, the joint space difference between the fixed image $F$ and the moving image $G$ can be described as follows:

\begin{equation}
\text{JSN}_{fg} = dy_0 - dy_1
\end{equation}

\subsubsection{Network architecture}
The architecture of the proposed registration network is illustrated in Fig.~\ref{fig:regiNet}. This network is based on residual convolution module \cite{he2016deep}, and it is implemented to obtain transformation parameters between the fixed image and the moving image.
The network takes the moving image, fixed image, and their segmentation masks as input, which are combined as four channels.
To reduce noise interference in the radiographic images of both fixed and moving images, a single-layer convolution is applied before entering the registration network.
Subsequently, after feature extraction by the registration network, eight output parameters are obtained through a full connection layer, these parameters include two sets of registration parameters containing the upper and lower bone regions of the joint, which are presented as significant outputs of the network. Then, these two sets of transformation parameters are applied to transform the moving image and its segmentation mask according to Eq.~\ref{eq:transform} and Eq.~\ref{eq:transSegmen}. Thus, the warped image can be obtained. In the training stage of the network, the loss function is the MSE of the distance between the fixed image and warped image, as defined in Eq.~\ref{eq:loss}.

\subsection{Implementation}
The joint segmentation and registration networks are trained and tested separately. The networks are implemented using Python language and Pytorch package \cite{paszke2019pytorch} on the workstation with single graphics processing unit (NVIDIA GeForce GTX TITAN V). The implementation details of networks are described as follows. 

\subsubsection{Segmentation network}
The loss function of segmentation network consisted of Sigmoid and binary cross entropy loss (BCELoss), and it was optimized using the RMSProp optimizer \cite{hinton2012neural}, with $\rho$ set to 0.9 and the epsilon set to 0.0001. The initial learning rate was 0.00001, and the training was carried out over 150 epochs with a batch size of 30. The dataset used for training contained 2577 images with corresponding labels that were expertly annotated. Additionally, the network was trained three times, and the final result was the average of the three runs to reduce the impact of random initialization.

\subsubsection{Registration network}
The network was trained using the Adam optimizer \cite{kingma2014adam}, with an initial learning rate of 0.001. ReduceLROnPlateau was used for adaptive learning rate reduction. The training process consisted of 500 epochs, with a batch size of 80. Furthermore, to minimize the impact of random initialization during training, the proposed registration network was trained three times, and the final result was obtained by averaging the three runs. The dataset used for training consisted of 1597 sets of moving and fixed images, along with their corresponding segmentation masks.

\begin{table}[!t]
\renewcommand{\arraystretch}{1}
\caption{Patient information in the clinical dataset}
\centering
\begin{threeparttable}
\setlength{\tabcolsep}{2.3mm}{
\begin{tabular}{p{3.5cm}<{\centering}p{2cm}<{\centering}p{2cm}<{\centering}}
\toprule
& Mean $\pm$ SD & Range\\
\midrule
Age at enrollment (year) & 56.11 $\pm$ 13.79 & 20.68 $\sim$ 88.00\\
Number of radiography & 4.30 $\pm$ 2.54 & 3 $\sim$ 17\\
Follow-up period (year) & 4.04 $\pm$ 3.44 & 0.88 $\sim$ 12.10\\
\bottomrule
\end{tabular}
}
\label{tab:clinicalInformation}
\end{threeparttable}
\end{table}

\begin{table}[!t]
\caption{Configuration parameters for radioactive imaging}
\label{tab2}
\centering
\begin{threeparttable}
\setlength{\tabcolsep}{0.8mm}{
\begin{tabular}{p{2.8cm}<{\centering}p{1.8cm}<{\centering}p{1.8cm}<{\centering}p{1.8cm}<{\centering}}
\toprule
& SARC & HMCRD & SCGH\\
\midrule
Model & DR-155HS2-5 & Radnext 32 & KXO-50G\\
Manufacturer & Hitachi & Hitachi & Toshiba \\
Aluminum filter (mm) & 1.5 & 0.5 & NO\\
Tube voltage (kV) & 42 & 50 & 45 \\
Tube current (mA) & 100 & 100 & 250 \\
Exposure time (mSec) & 20 & 25 & 14 \\
Source to image (cm) & 100 & 100 & 100\\
Resolution (mm/pixel) & 0.175 & 0.15 & 0.15\\
Image size (pixel) & 2010$\times$1490 & 2010$\times$1490 & 2010$\times$1490\\
Bit depth (bit) & 12 & 10 & 10\\
\bottomrule
\end{tabular}
}
\begin{tablenotes}
\item[*] \textbf{SARC}: Sagawa Akira Rheumatology Clinic. \\\textbf{HMCRD}: Hokkaido Medical Center for Rheumatic Diseases. \\\textbf{SCGH}: Sapporo City General Hospital. 
\end{tablenotes}
\end{threeparttable}
\end{table}

\begin{figure*}[!t]
\includegraphics[width=\textwidth]{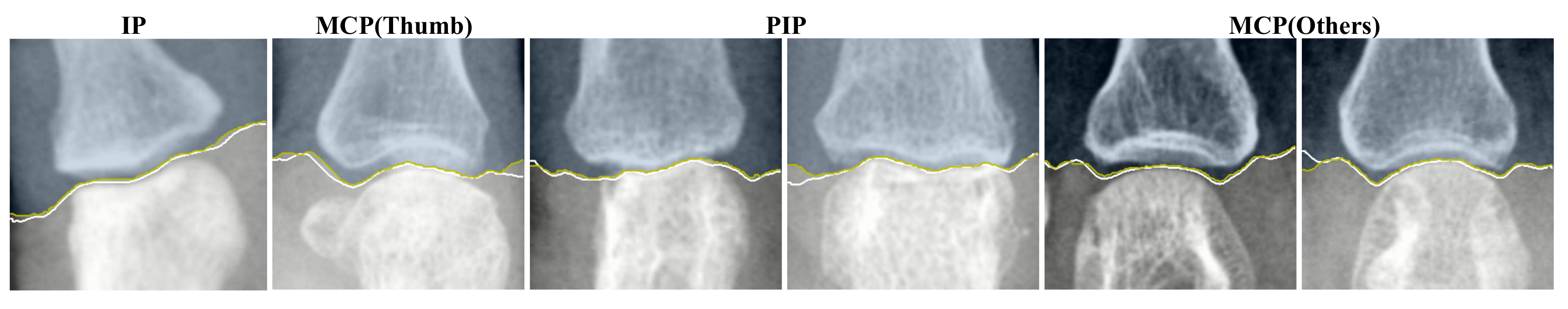}
\caption{Experiments of the proposed segmentation network. White lines represent the manual label of segmentation, and yellow lines represent the predicted segmentation by using the network.}
\label{fig:segeRes}
\end{figure*}

\subsection{Dataset}
To evaluate the performance of our network, a clinical dataset was prepared. Our study was conducted in accordance with the guidelines of the Declaration of Helsinki and approved by the Ethics Committee of the Faculty of Health Sciences, Hokkaido University (approval number: 19 - 46).
The dataset used in this study consisting of 675 hand posteroanterior projection (PA) radiographs from 80 RA patients, of which 88.5\% were female, and detailed patient information is summarized in Table~\ref{tab:clinicalInformation}. Images are from three institutions, \textit{Sagawa Akira Rheumatology Clinic} (Sapporo, Japan), \textit{Hokkaido Medical Center for Rheumatic Diseases} (Sapporo, Japan), and \textit{Sapporo City General Hospital} (Sapporo, Japan), they own separate X-ray systems.
The dataset is managed using digital imaging and communications in medicine (DICOM) standard. For a detailed description of the imaging parameters, please refer to Table~\ref{tab2}.

Finger joint images were extracted from the hand images using the finger joint detection method described in \cite{ou2022sub}.
The finger joints were scored by a radiologist following extensive training, and only early-stage RA cases with a SvdH hand score of 0 were included, as narrowed joint space can affect segmentation accuracy and bone erosion can cause damage to the bone margin in advanced RA.
In addition, we retained only images from patients who underwent radiography of the hand at least three times to enable the calculation of standard deviation. Table~\ref{tab1} shows the distribution of data for different joints.

For the segmentation task, we divided the dataset into an 80$\%$ training set (4854 finger joint images) and 20$\%$ for testing (1213 finger joint images). 
For the registration task, since registration data requires paired images (fixed and moving images), the dataset was produced as follows: for each joint with multiple images, we constructed the Cartesian set of each joint image set as the overall dataset. 
We selected the following data as the training set (1597 finger joint image pairs): using the middle index of the sequence as the segmentation point, a pair of images matched one-to-one by images smaller than the middle index and larger than the middle index. Relatively, other data were used as test data (3604 finger joint image pairs).

\begin{table}[!t]
\caption{The number of finger joint images in our clinical dataset}
\label{tab1}
\centering
\begin{threeparttable}
\setlength{\tabcolsep}{5.4mm}{
\begin{tabular}{lcccc}
\toprule
&IP&PIP&MCP&OverAll\\
\midrule
Thumb & 561 & N/A  & 569 & \\
Index & N/A & 636  & 672 & \\
Middle & N/A & 647  & 599 & \\
Ring & N/A & 514 & 626 & \\
Small & N/A & 560 & 683 & \\
\textbf{Overall} & \textbf{561} & \textbf{2357} & \textbf{3149} & \textbf{6067}\\
\bottomrule
\end{tabular}
}
\begin{tablenotes}
\item[*] \textbf{IP}: Interphalangeal joint. \quad\textbf{PIP}: Proximal interphalangeal joint. \\\textbf{MCP}: Metacarpophalangeal joint. 
\end{tablenotes}
\end{threeparttable}
\end{table}

\section{Experiments and discussion}
\label{sec:exp}

\subsection{Segmentation experiments}
\subsubsection{Segmentation evaluation}
With manual annotations as ground truth, the segmentation performance of U-net++ in this work was quantitatively evaluated with the following six metrics \cite{zhang1996survey}: (1) mIoU, (2) Sensitivity(SEN) (3) Specificity (SPC), (4) Dice Similarity Coefficient (DSC), (5) accuracy (ACC), which are defined as follow:
\begin{itemize}
\item \textbf{mIoU}: The ratio of the intersection and the concatenation of the two sets of Ground Truth and predicted result.
\item \textbf{SEN}: The percentage of Ground Truth that is correctly segmented.
\item \textbf{SPC}: The percentage of non-Ground Truth regions that are correctly segmented.
\item \textbf{DSC}: Similarity between Prediction and Ground Truth.
\item \textbf{ACC}: The percentage of correctly predicted pixels to the total pixels.
\end{itemize}

\begin{table}[!t]
\caption{The performance in different evaluation metrics}
\label{tab:segeRes}
\centering
\setlength{\tabcolsep}{2.8mm}{
\begin{tabular}{cccccc}
\toprule
&mIoU&SEN&SPC&DSC&ACC\\
\midrule
IP & 0.95900 & 0.97782  & 0.98292 & 0.97896 & 0.98028\\
PIP & 0.96227 & 0.97115  & 0.99191& 0.98056  & 0.98166\\
MCP & 0.95396 & 0.96422  & 0.99094& 0.97615  & 0.97821\\
\textbf{Overall} & \textbf{0.95779} & \textbf{0.96835}  & \textbf{0.99052} & \textbf{0.97819}  & \textbf{0.97980}\\
\bottomrule
\end{tabular}
}
\end{table}

\begin{figure*}
\includegraphics[width=\textwidth]{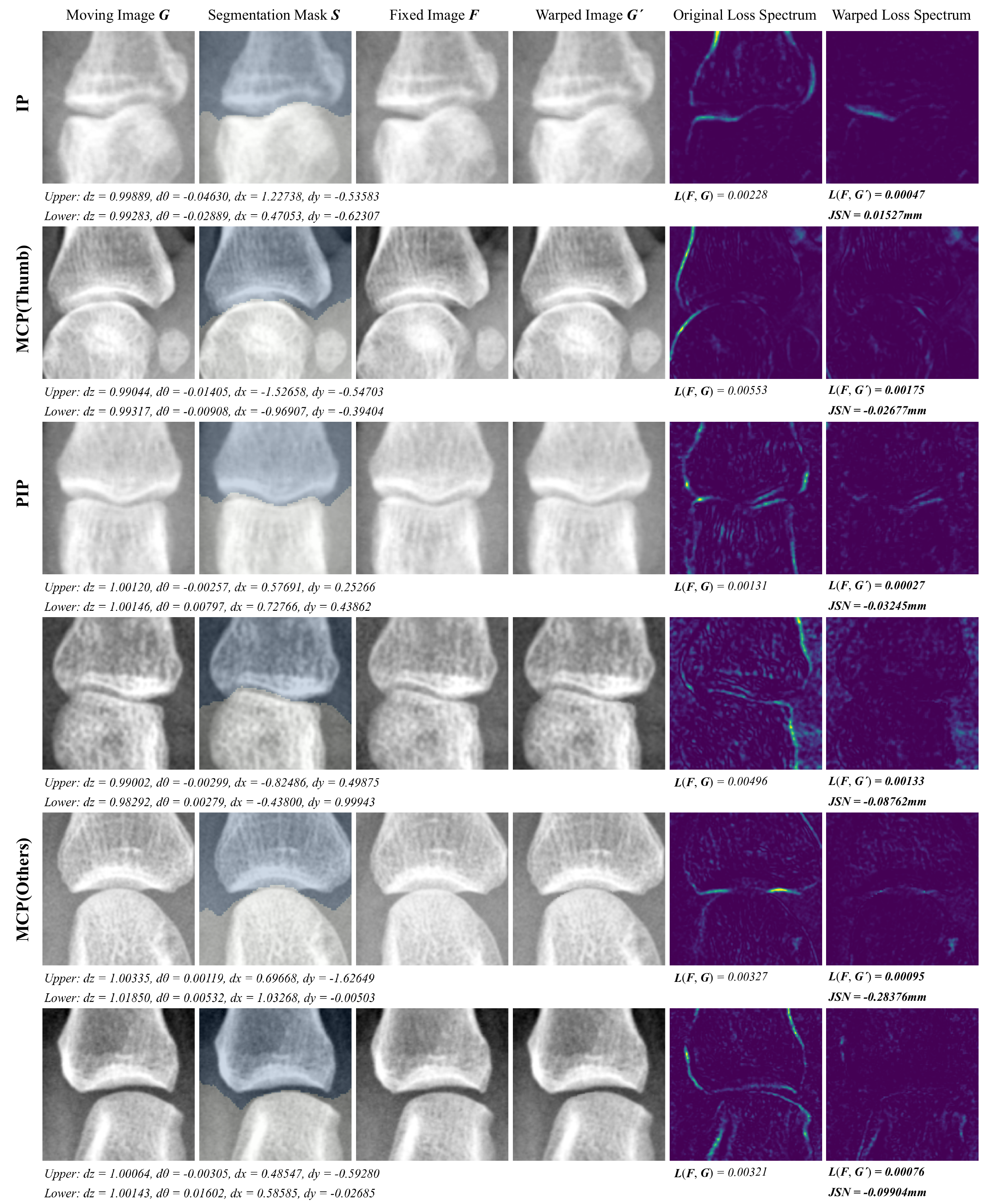}
\caption{Experiments of the proposed registration network. Considering that the MCP joint of the thumb are partially different from it of the other fingers, we have divided the finger joints into four categories here; IP, MCP (thumb), PIP and MCP (others). Left three columns are inputs, the moving image $G$, the segmentation mask $S$ and the fixed image $F$. The fourth column is the warped image $G^{\prime}$. Right two columns are Euclidean distance loss spectrums, the original loss spectrum $E_{FG}$ and the warped loss spectrum $E_{FG^{\prime}}$. The quantified transformation parameters $P_0$, $P_1$, the original loss $L(F,G)$ and the warped loss $L(F,G^{\prime})$ are listed below the images.}
\label{fig:regiRes}
\end{figure*}

\subsubsection{Segmentation results}
The segmentation results on the dataset with 1104 finger joint images were obtained by our segmentation method. Different evaluation metrics were used to evaluate the IP, PIP, and MCP joints separately, and the results are shown in Fig.~\ref{fig:segeRes} and Table~\ref{tab:segeRes}. The visualization and evaluation results imply that the automatic segmentation results obtained by our method are accurate and sensitive compared with the ground truth.

\begin{figure*}
\includegraphics[width=\textwidth]{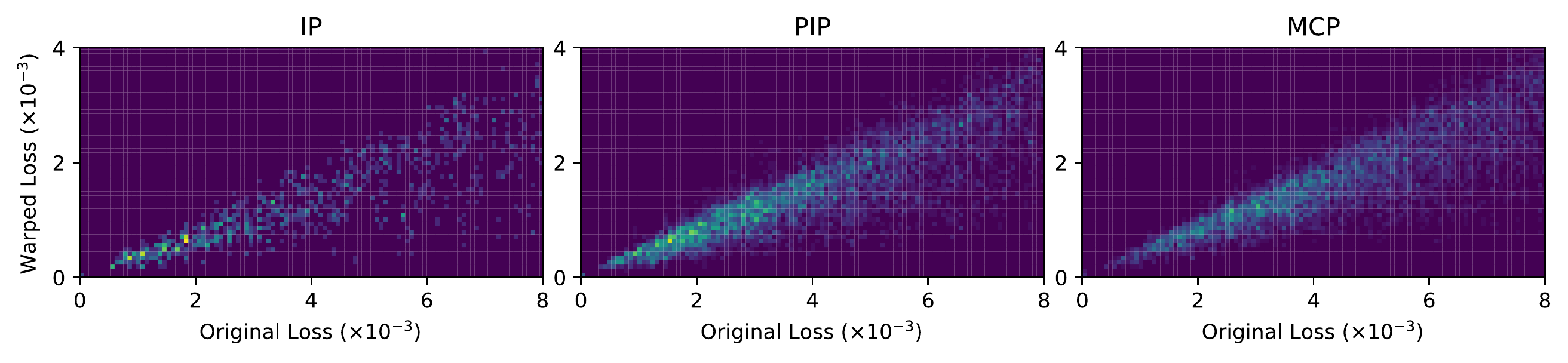}
\caption{The heat maps of the original loss $L(F,G)$ and the warped loss $L(F,G^{\prime})$. This set of heat maps exhibited the distribution and relationship between the original loss and the warped loss. }
\label{fig:lossRes}
\end{figure*}

\subsection{Registration experiments}

\subsubsection{Registration evaluation}
There is evidence that visual measurement also can be greatly affected by noise in radiographic images \cite{ou2022sub}. Their phantom experiments indicate that the manually annotated data can have sub-pixel mean error, which may result in sub-pixel deviation in the algorithm evaluation. Thus, metrics that do not rely on manually annotated data are used for algorithm performance evaluation.
The experiments of our registration network are summarized in Table~\ref{tab:registration} by using four metrics; the standard deviation $\sigma$ defined in \cite{ou2019automatic}, the standard deviation $\sigma^{\prime}$ defined in \cite{langs2008automatic}, the mismatching ratio and the warped loss.

\begin{itemize}
\item \textbf{Standard deviation $\sigma$}: The metric used to evaluate the accuracy of JSN progression quantification.
\item \textbf{Standard deviation $\sigma^{\prime}$}: The metric used to demonstrate uncertainty.
\item \textbf{Mismatching ratio}: The percentage of mismatching case, which can measure the robustness.
\item \textbf{Warped loss}: The difference between the warped image and the fixed image, which can quantify the performance of the registration network.
\end{itemize}

The standard deviation $\sigma$ of multiple measurements are computed to demonstrate the reliability of registration network without the ground truth \cite{ou2022sub}. The definition of standard deviation can be described using the following instance. In case of three images $F, G$ and $K$, the $JSN_{FG-I}$ between image $F$ and image $G$ can be indirectly calculated by introducing intermediate image $I$, as given as follow:
\begin{equation}
JSN_{FG-I} = JSN_{FI} + JSN_{IG}
\end{equation}
Considering a set of images, the $\overline{JSN_{fg}}$ can be obtained by taking the average of multiple measurements.
\begin{equation}
\overline{JSN_{FG}} = \frac{1}{n} {\sum_{I=1}^{n}} JSN_{FG-I}
\end{equation}
Therefore, the standard deviation $\sigma_{fg}$ of $JSN_{fg}$ can be defined as follow:
\begin{equation}
\sigma_{FG} = \sqrt{\frac{1}{n}\left({\sum_{I=1}^{n}}JSN_{FG-I}-\overline{JSN_{FG}}\right)^{2}}
\label{eq:sd1}
\end{equation}

We utilize the standard deviation introduced in \cite{langs2008automatic} to demonstrate the reliability of our method. In case of two images $F$ and $G$, the image $G_j$ can be obtained by adding a random translation (-3, +3 pixel) in image $G$ along the x-axis and y-axis. The standard deviation $\sigma_{FG}^{\prime}$ of images $F$ and $G$ can be defined by using the $JSN_{FG-j}$ between image $F$ and image $G_i$. 

\begin{equation}
\overline{JSN_{FG}} = \frac{1}{10}{\sum_{j=1}^{10}}JSN_{FG-j}
\end{equation}
\begin{equation}
\sigma_{FG}^{\prime} = \sqrt{\frac{1}{10}\left({\sum_{j=1}^{10}} JSN_{FG-j}-\overline{JSN_{FG}}\right)^{2}} 
\label{eq:sd2}
\end{equation}

When calculating the standard deviation $\sigma_{FG}^{\prime}$, outliers will be removed as mismatches.


\begin{figure}
\includegraphics[width=\linewidth]{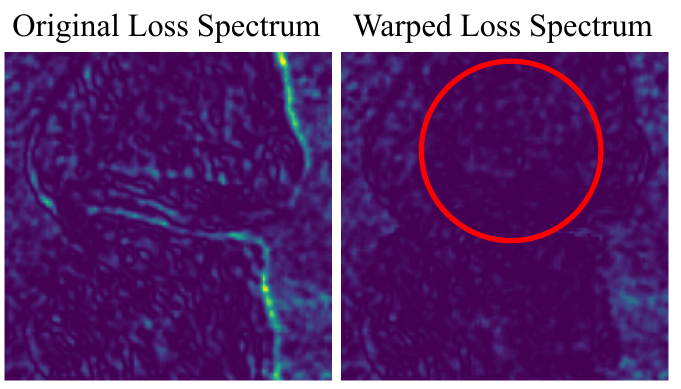}
\caption{The bony texture varies as the bone thickness/diameter vary (in response to muscle activity or weight) or due to any changes in the imaging angle. Therefore, it is difficult to reduce the loss on the bone surface region to 0, as shown in the highlighted region. These irregular variation on bony texture is also a major part of the warped loss.}
\label{fig:surfaceLoss}
\end{figure}

\begin{figure*}
\includegraphics[width=\textwidth]{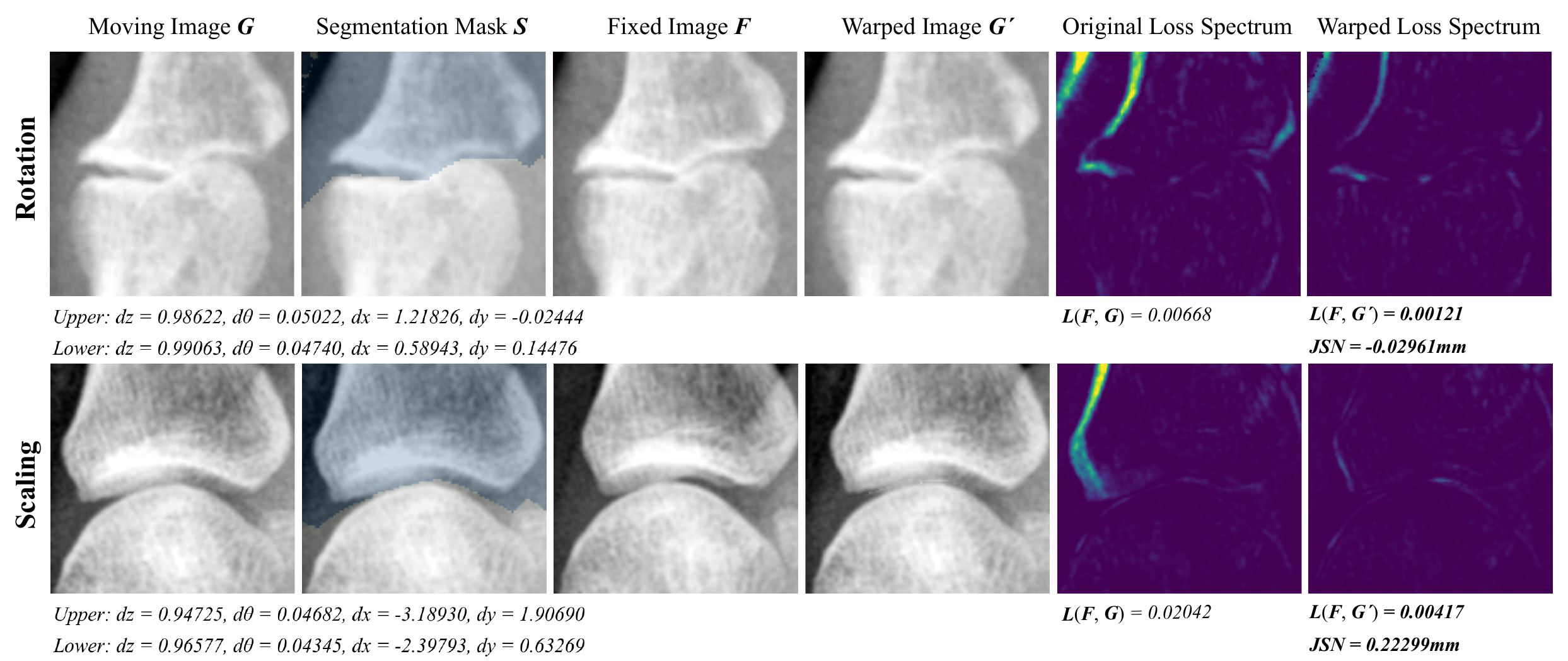}
\caption{Experiments of the proposed registration network in specific cases. Those figures show the robustness of the proposed registration network to rotation, scaling, and bone erosion.}
\label{fig:outRes}
\end{figure*}

\begin{table*}[!t]
\caption{The mean standard deviation in millimeter, the mismatching ratios and the warped Euclidean distance loss for our image registration network}
\label{tab:registration}
\centering
\setlength{\tabcolsep}{1.6mm}{
\begin{tabular}{p{1cm}<{\centering}p{1cm}<{\centering}p{1cm}<{\centering}p{1cm}<{\centering}p{0cm}<{\centering}p{1cm}<{\centering}p{1cm}<{\centering}p{1cm}<{\centering}p{0cm}<{\centering}p{1cm}<{\centering}p{1cm}<{\centering}p{1cm}<{\centering}p{0cm}<{\centering}p{1cm}<{\centering}p{1cm}<{\centering}p{1cm}<{\centering}}
\toprule
& \multicolumn{3}{c}{Standard Deviation $\sigma$} & & \multicolumn{3}{c}{Standard Deviation $\sigma^{\prime}$} & & \multicolumn{3}{c}{Mismatching Ratio (\%)} & & \multicolumn{3}{c}{Warped Loss}\\
\cline{2-4}\cline{6-8}\cline{10-12}\cline{14-16}
&IP&PIP&MCP& &IP&PIP&MCP& &IP&PIP&MCP& &IP&PIP&MCP\\
\midrule
Thumb & 0.0864  & N/A  & 0.0688  &  
			& 0.0415    & N/A  & 0.0394  & 
			& 3.31 & N/A & 0.53 &  
			& 0.0049  & N/A  & 0.0031 \\
Index & N/A & 0.0496  &  0.0689  &  
		  & N/A & 0.0354  &  0.0373  & 
		  & N/A & 1.41 & 1.07 &  
		  & N/A & 0.0031   & 0.0029 \\
Middle & N/A & 0.0491  & 0.0571 &  
			& N/A & 0.0400  & 0.0331 & 
			& N/A & 0.65 & 0.01 &  
			& N/A & 0.0021 & 0.0028 \\
Ring & N/A & 0.0423   & 0.0610   &  
		 & N/A & 0.0358   & 0.0313  & 
		 & N/A & 0.14 & 0.44 &  
		 & N/A & 0.0020   & 0.0031 \\
Small & N/A & 0.0410  & 0.0745 &  
		  & N/A & 0.0294  & 0.0312 & 
		  & N/A & 0.54 & 0.34 &  
		  & N/A & 0.0023 & 0.0035\\
\textbf{Overall} & \textbf{0.0864}  & \textbf{0.0455}  & \textbf{0.0661} &  & \textbf{0.0415} & \textbf{0.0351} & \textbf{0.0345} & & \textbf{3.31} & \textbf{0.68} & \textbf{0.48} &  & \textbf{0.0049} & \textbf{0.0024} & \textbf{0.0031}\\
\bottomrule
\end{tabular}
}
\end{table*}

\subsubsection{Registration results}
The experimental results of our registration network over various finger joints are shown in Fig.~\ref{fig:regiRes}.
In this figure we have divided the finger joints into four categories: IP, MCP (thumb), PIP and MCP (others). 
In the original loss spectrums and the warped loss spectrums, the highlighted regions represent relative displacement between fixed image and moving/warped image.
As shown in Fig.~\ref{fig:regiRes}, the proposed registration network can effectively reduce the highlight regions, especially around bone margin region.
Considering that the bone margin information is used to determine JSW or JSN progression in clinical data, rather than bony texture information.
Moreover, the bony texture will vary as the bone thickness vary, or if there is any change of the imaging angle.
Therefore, the loss around bone margin region is more important than the loss of bony texture.
These irregular variation on bony texture is also the primary cause of loss in rigid registration network, as shown in Fig.~\ref{fig:surfaceLoss}.
This demonstrates that the proposed registration network can accurately quantify the transformation parameters between the fixed image and moving image.

Our experiments show that the standard deviation $\sigma$, the mismatching ratio, and the mean original loss of IP joint images are much higher than others. 
There are substantial evidence that thumb movements are more independent compared to other fingers \cite{ingram2008statistics}.
As a result, the IP joint exhibits distinct characteristics when the hand posture is altered.
Inconsistent hand posture is primarily shown radiographically as rotation or scaling in PIP and MCP joints.
However, the rotation of the thumb MCP joint can lead to rolling in the thumb, thereby altering the projection angle of the IP joint.
This can significantly affect the IP joint characteristics, leading to decreased quantification accuracy and potential mismatches.
Any further improvement in the algorithms’ robustness and reduced mismatches on IP joints continues to be a challenge. 

Figure~\ref{fig:lossRes} demonstrates the distribution and the relationship between original loss and warped loss in various joint images.
Because the characteristics of various joints are different, and the distribution is not the same.
For the reason we discussed above, the mean and variance of the original loss and the warped loss of IP joint images are much higher than others.
In all three kinds of finger joint images, our proposed registration network can effectively control the loss.
In 94.6\% of the registration cases, the warped loss is less than half compared to the original loss. 

Actually, the warped loss is difficult to decrease infinitely in the rigid registration, because the bone features across multiple radiographic images can be different, including bony texture, margin information from finger bending, and bone erosion. 
These variations pose significant obstacles to achieve successful rigid registration.
Moreover, experiments involving these factors typically yield high original loss, leading to a higher occurrence of mismatch cases in the high original loss region.
In contrast, our proposed method, as shown in Fig.\ref{fig:lossRes}, demonstrates high robustness and a low mismatching ratio in this region. 
Table\ref{tab:registration} further highlights the effectiveness of our method in controlling the mismatching ratio of various finger joint images, particularly PIP and MCP joint images.

\begin{table*}[!t]
\caption{Comparison with related works. The mean standard deviation in millimeter and the mismatching ratios for respective joints.}
\label{tab:compare}
\centering
\begin{threeparttable}
\setlength{\tabcolsep}{1mm}{
\begin{tabular}{rlcccccccccccccccccc}
\toprule
& \multirow{2}{*}{Method$^\dag$} & Dataset & Resolution & Output & \multicolumn{4}{c}{Standard Deviation $\sigma$} &  & \multicolumn{4}{c}{Standard Deviation $\sigma^{\prime}$} &  & \multicolumn{4}{c}{Mismatching Ratio (\%)}\\
\cline{6-9} \cline{11-14} \cline{16-19}
& & (Images)&(mm/pixel)& Metric &IP&PIP&MCP&OverAll&  &IP&PIP&MCP&OverAll&  &IP&PIP&MCP&OverAll\\
\midrule
TMI$^\prime$08\cite{langs2008automatic} & ASM & 160 MCP & 0.0846 & JSW & - & - & - & - &  
            & - & - & 0.0800 & 0.0800 &  
            & - & - & - & - \\
JBHI$^\prime$23\cite{ou2022sub} & PIPOC & 549 & 0.175 & JSN 
        & 0.093 & 0.053 & 0.050 & 0.056 &  
        & - & 0.0095 & 0.0061 & 0.0076 &  
        & 7.2 & 3.5 & 2.8 & 3.5 \\
\textbf{This work} & \textbf{CNN} & \textbf{675} & \textbf{0.175/0.15} & \textbf{JSN} 
        & \textbf{0.086} & \textbf{0.046} & \textbf{0.066} & \textbf{0.066} &  
        & \textbf{0.0415} & \textbf{0.0351} & \textbf{0.0345} & \textbf{0.0370} &  
        & \textbf{3.31} & \textbf{0.68} & \textbf{0.48} & \textbf{1.49}\\
\bottomrule
\end{tabular}
}
\begin{tablenotes}
\item[$\dag$] \textbf{ASM}: Active Shape Models. \quad \textbf{PIPOC}: Partial image phase only correlation. \quad \textbf{CNN}: Convolutional Neural Network.
\end{tablenotes}
\end{threeparttable}
\end{table*}

\subsubsection{Comparison with related works}
Consider that during clinical radiographic imaging, inconsistent hand posture of the patient or different imaging equipment may be reflected in the radiography as rotation or scaling of the bones. Also, as the RA progresses, bone erosion gradually destroys the margin information of bones. These changes in bone margin information due to rotation, scaling, or bone erosion pose a challenge for rigid image registration based JSN progression quantification in RA.

According to experiments of PIPOC based JSN progression quantification in \cite{ou2022sub}, PIPOC is susceptible to noise, rotation, and scaling, thus, inconsistent hand posture is the major mismatch reason in PIPOC. This can be broadly divided into two cases. (i) The inconsistent angle between the upper and lower bones of the joint, as show in the first row of Fig.~\ref{fig:outRes}. This inconsistent joint angle is shown radiographically as a rotation. (ii) The bending of the fingers, as show in the second row of Fig.~\ref{fig:outRes}. In this case, there will be obvious scale differences between the upper and lower bones. In the paper \cite{ou2022sub}, it is reported that due to the characteristics and limitation of PIPOC, the inconsistency of the angles or scales of upper and lower bones can easily cause mismatches. As shown in the warped loss spectrums of Fig.~\ref{fig:outRes}, the registration network proposed in this work can improve the robustness for rotation and scaling, and it can accurately quantify the angle and scale difference. This improvement can effectively reduce the mismatching ratios when compared to PIPOC, as shown in Table~\ref{tab:compare}.


Table~\ref{tab:compare} summarizes the comparison with other joint space quantification work in RA. 
We can observe that the image registration based JSN progression quantification framework can achieve lower standard deviation $\sigma^{\prime}$ when compared to the margin detection based JSW quantification framework. This shows that the image registration based framework has lower uncertainty and has greater potential for accuracy and sensitivity.
Neural network is a non-linear function \cite{gawlikowski2021survey}, which can lead to higher uncertainty when compared to traditional image processing algorithms. The major uncertainty in this work is aleatoric uncertainty, that exist due to noise, and is irreducible by improving the quality or quantity of data \cite{indrayan2017medical, mehrtash2020confidence}. As the standard deviations $\sigma$ shown in the Table~\ref{tab:compare}, the impact of aleatoric uncertainty can be controlled, and can attain the standard deviation $\sigma$ close to the PIPOC based JSN progression quantification work in \cite{ou2022sub}. This demonstrates that our proposed network can achieve lower mismatching ratio while ensuring similar accuracy to previous studies. 

\section{Conclusion and Future work}
\label{sec:conl}
In this work, we propose a deep learning method for joint space narrowing (JSN) progression quantification in rheumatoid arthritis (RA). The proposed method includes an image segmentation network based on U-net++, and a ResNet-like deep registration network for displacement quantification. Our extensive clinical experiments demonstrate that this work can achieve sub-pixel level accuracy monitoring of joint space in the early stage of RA.

Medical image analysis for RA can be classified into three groups: (i) joint space width (JSW) quantification based on margin detection, (ii) SvdH scoring based on machine learning classification, (iii) JSN progression quantification based on image registration. And the method in (iii) is used in our work. Compared to two other mainstream frameworks, the image registration based JSN progression quantification framework is highly advantageous in monitoring and drug management of RA due to its superior accuracy and sensitivity.

In their work, \cite{ou2022sub} proposed a novel image registration algorithm called partial phase only correlation (PIPOC) for JSN progression quantification, achieving sub-pixel accuracy in both phantom and clinical experiments.
This has significant clinical implications for closely monitoring the condition of RA and providing evidence for drug management.
However, PIPOC has limitations as it can only quantify bone displacement on the x-axis and y-axis and is susceptible to noise, rotation, scaling, and even cause mismatches. To overcome these limitations, we propose an intra-subject rigid registration network that can simultaneously quantify four transformation parameters (scaling, rotation, x-axis and y-axis displacements), thereby improving the robustness to noise, scaling, and rotation.
Our approach can handle complex clinical situations and reduce mismatches due to inconsistent angle and spatial resolution of radiography images. Additionally, our approach provides a visualization loss as a reliability indicator that can be used by radiologists and rheumatologists to assess the quantification reliability, thus, making it a promising tool for future clinical applications.

Recently, non-rigid registration network based on deformation fields have received sufficient attention and development. 
Our algorithmic process is a kind of regional image registration.
An interesting direction for future research could be the incorporation of segmentation information to immobilize the target region of the deformation field. This approach draws on the advantages of the deformation field, enabling the quantification of JSN progression in complex joint regions, such as wrist joints. This could lead to more comprehensive monitoring in the early stages of RA and provide novel ideas for registration-based joint space measurements.
 
\section{Acknowledgments}
The authors would like to sincerely thank Akira Sagawa, MD, PhD, Sagawa Akira Rheumatology Clinic (Sapporo, Japan), Masaya Mukai, MD, PhD, Sapporo City General Hospital (Sapporo, Japan) and kazuhide Tanimura, MD, Hokkaido Medical Center for Rheumatic Diseases (Sapporo, Japan) for image data preparation.

\bibliographystyle{IEEEtran}
\bibliography{reference}

\end{document}